\documentclass[12pt,preprint]{aastex}

\newcommand\OUCHII{UCH\,{\sc ii}}
\newcommand\UCHII{UCH}
\newcommand\HII{H\,{\sc ii}}

\newcommand\HI{H\,{\sc i}}

\newcommand\kms{km~s$^{-1}$}

\newcommand\cmthree{cm$^{-3}~$}

\newcommand\etal{et al.}
\newcommand\be{\begin{equation}}
\newcommand\ee{\end{equation}}
\newcommand\bea{\begin{eqnarray}}
\newcommand\eea{\end{eqnarray}}

\newcommand\alfven{Alf$\acute{v}$en~}

\catcode`\@=11
\newcommand{\gsim}{${\mathrel{\mathpalette\@versim>}}$}
\newcommand{\lsim}{${\mathrel{\mathpalette\@versim<}}$}
\newcommand{\@versim}[2]{\lower 2.9truept \vbox{\baselineskip 0pt \lineskip
    0.5truept \ialign{$\m@th#1\hfil##\hfil$\crcr#2\crcr\sim\crcr}}}
\catcode`\@=12

\slugcomment{To appear in ApJ Letters}

\shorttitle{Magnetic fields near \OUCHII\ regions}
\shortauthors{D. Anish Roshi}

\begin{document}

\title{Magnetic fields at the periphery of \OUCHII\ regions from carbon recombination line observations}

\author{D. Anish Roshi\altaffilmark{1}}
\altaffiltext{1}{Raman Research Institute, Sadashivanagar, Bangalore 560 080, India;
anish@rri.res.in }

\begin{abstract}
Several indirect evidences indicate a magnetic origin for the
non-thermal width of spectral lines observed toward molecular
clouds. In this letter, I suggest that the origin of the non-thermal width of 
carbon recombination lines (CRLs) observed from photo-dissociation
regions (PDRs) near ultra-compact \HII\ regions is magnetic
and that the magnitude of the line width is an estimate of the \alfven speed. 
The magnetic field strengths estimated based on this suggestion
compare well with those measured toward molecular clouds with
densities similar to PDR densities. 
I conclude that multi-frequency CRL observations
have the potential to form a new tool to determine 
the field strength near star forming regions. 
\end{abstract} 


\section{Introduction}
\label{sec1}

Observations of high-density molecular line tracers toward ultra-compact
\HII\ regions (\UCHII s) reveal that these \HII\ regions are embedded
in dense (\gsim $10^5$ \cmthree) molecular clouds 
(eg. \nocite{cetal90}Churchwell, Walmsley \& Cesaroni 1990; 
\nocite{c02}Churchwell 2002 and references there in). 
Far ultra-violet (FUV; 6 to 13.6 eV) radiation from massive stars 
within the \UCHII s 
heats the dense molecular material in this interface, 
producing a Photo Dissociation Region (PDR) 
(see \nocite{ht97} Hollenbach \& Tielens 1997).
Carbon recombination lines (CRLs) from such
regions have been detected toward a large number
of \UCHII s, establishing the presence of 
dense PDRs near most \UCHII s
(\nocite{retal05a}Roshi \etal\ 2005a).

\begin{figure}
\includegraphics[width=3.0in, height=3.3in, angle = -90]{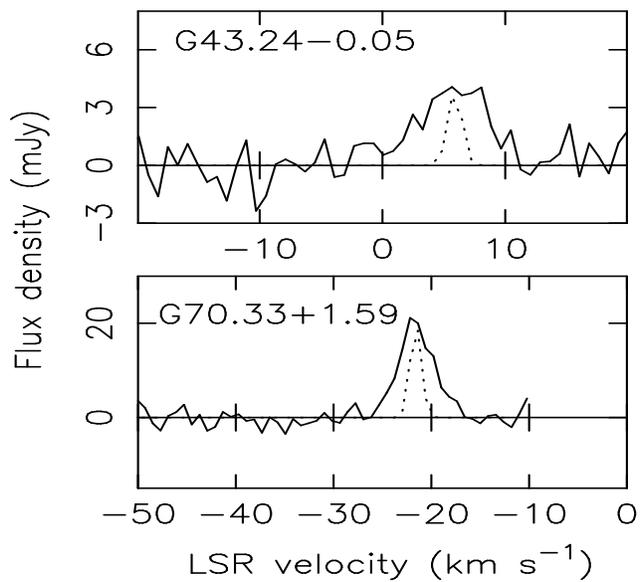}
\caption{Examples of carbon recombination line spectra toward 
\UCHII s G43.24$-$0.05 and G70.33$+$1.59 observed near 9 GHz
(Roshi \etal\ 2005a). The Gaussian profiles plotted
in dotted line correspond to an assumed PDR gas temperature of 1000 K.
The large difference between the observed line profiles and the Gaussian
curves demonstrates the dominance of non-thermal motions in the
PDR. The LSR velocity is with respect to the C89$\alpha$ 
(9.1779 GHz) transition.}
\label{fig1}
\end{figure}

The observed width of the CRLs from PDRs associated with
\UCHII s is typically between 4 and 8 \kms. The gas temperature
in the PDR, obtained by non-LTE modeling of CRL emission at
multiple frequencies, is in the range 200 to 1000 K 
(\nocite{retal05b}Roshi \etal\ 2005b;
\nocite{getal98}Garay \etal\ 1998; 
\nocite{nwt94} Natta, Walmsley \& Tielens 1994). 
Thus the expected thermal contribution to the line width
is at least a factor of two smaller than the observed width
(see Fig.~\ref{fig1}). This larger observed width indicates that 
the CRL width is dominated by
supersonic motions. In fact, a similar situation exists
in molecular clouds. It has long been recognized that
the observed width of spectral lines from molecular clouds 
have a non-thermal component and this component is often supersonic 
(eg. \nocite{bmw64}Barrett, Meeks and Weinreb 1964). The origin
of the non-thermal component of the line width is now 
attributed to either ``turbulence'' (\nocite{metal74}Morris \etal\ 1974; 
\nocite{ze74}Zuckerman \& Evans 1974) or \alfven waves
(\nocite{am75}Arons \& Max 1975; \nocite{m75}Mouschovias 1975; 
see also \nocite{sal87}Shu, Adams \& Lizano 1987 and references there in). 

Detailed theoretical models for both ``turbulence'' and
\alfven waves in molecular clouds are yet to be worked out. 
But ``turbulence'' without any magnetic field
is a less plausible proposition since observations demand supersonic
turbulence and the rapid dissipation in shocks cannot
sustain such turbulent motions to time scales larger than the free-fall 
time (eg. \nocite{sal87}Shu \etal\ 1987). The
decay of \alfven waves are slow compared to supersonic turbulence
though simulations show that they may also need a driving mechanism
(\nocite{m03}MacLow 2003).
Thus \alfven waves appear to be a more tenable explanation for
the observed line widths
(\nocite{am75}Arons \& Max 1975; \nocite{m75}Mouschovias 1975).

Observational data also seem to indirectly support the magnetic origin
of the line width. Analysis of the width of spectral lines observed 
toward molecular clouds shows power-law relationships between 
the non-thermal component of the velocity dispersion, the 
cloud size, density and the magnetic field strength 
(\nocite{l81}Larson 1981; \nocite{th86}Troland \& Heiles 1986;
 \nocite{mg88a}Myers \& Goodman 1988a, 1988b;
\nocite{mp95}Mouschovias \& Psaltis 1995; \nocite{c99}Crutcher 1999). 
These relationships
are expected in self-gravitating, magnetically supported clouds
since the \alfven waves in such clouds affect the
observed line width (\nocite{m87}Mouschovias 1987; Myers \&
Goodman 1988a, 1998b). 

In this letter, I use the observed non-thermal widths of CRLs
to estimate the magnetic field strength in PDRs near UCHs.
The field strength is estimated by assuming that the
non-thermal velocity dispersion of carbon lines give an estimate of
the \alfven speed in the PDR. This assumption is 
based on a comparative study of the characteristics exhibited
by the CRL and molecular line data. The data and the comparative
study are presented in Sections~\ref{sec2} and \ref{sec3}
respectively.  The magnetic field strength
is obtained by combining the velocity dispersion of CRL
and the density derived from modeling carbon line emission 
(see Section~\ref{sec4}).
To our knowledge the only attempt to compute the magnetic
field using CRLs was made by \nocite{v89}Vallee (1989), where
he had applied a model for shocks at the edge of molecular
clouds to deduce the field strength.
 
\section{Carbon recombination line data}
\label{sec2}

For the present analysis I use CRL data toward
\UCHII s since the PDR thickness is small in these cases and therefore
large scale motions (such as outflows) may not affect the line
width (see Section~\ref{sec5}). Table.~\ref{tab1} summarizes the 
CRL data and results obtained from modeling this data toward 14 \UCHII s. 
Listed are the source name,
observed CRL transition, FWHM of CRLs\footnote{The relationship
$\Delta V = \sqrt{8ln(2)}~\sigma_{NT}$ is used in this paper to relate the
FWHM line width $\Delta V$ to the dispersion $\sigma_{NT}$ for a Gaussian
line model}, PDR gas temperature, carbon ion density, 
number density of hydrogen atoms in the PDR, line-of-sight extent of
the PDR ($L_{\parallel}$) and the estimated magnetic field
strength (see Section~\ref{sec4}). 

The quality of the CRL data obtained toward all 14 sources is good;
the uncertainties are mostly in the parameters derived by modeling
line emission.
The frequencies of the selected CRLs are below $\sim$ 15 GHz. As
inferred from modeling, carbon line emission is dominated by 
stimulated emission at these frequencies (\nocite{nwt94} Natta \etal\ 1994;
\nocite{retal05a}Roshi \etal\ 2005a).
The data toward the first 7 sources in Table~\ref{tab1} were obtained
using interferometric observations. 
The line widths for the first six sources were obtained from 
profiles averaged over regions where lines were detected. The
properties of the PDR toward W3A were not well constrained; I
have taken representative model parameters from 
\nocite{kag98}Kantharia, Anantharamaiah \& Goss (1998). 
Data toward the remaining 7 sources were obtained with the Arecibo telescope. 
Modeling of CRL emission toward these sources was limited by: 
(a)poor angular resolution, (b)flux density calibration error and 
(c) detection of line only near 9 GHz. 
Thus modeling this data set has given only
constraints on the physical properties of the PDR. 
The PDR gas temperature of 500 K for the Arecibo sources 
and W48A, W49G and W49J listed in Table~\ref{tab1} is a representative value. 
The carbon ion density and $L_{\parallel}$ 
obtained from modeling depend non-linearly on the gas temperature.
Converting the ion density to neutral density is somewhat uncertain
due to the unknown depletion factor and the fraction of molecular
hydrogen in the CRL forming region. Based on the examination of
the models for W48A and our experience in modeling other CRL
data set (\nocite{retal05a}Roshi \etal\ 2005; 
\nocite{retal06}Roshi \etal\ 2006),
it is expected that the values for temperature in Table~\ref{tab1} 
can vary by a factor of 2 and the values for 
neutral density and $L_{\parallel}$ 
can vary by a factor of 4. The model parameters are derived using
a simplified geometry of plane parallel PDR slabs.
A comparison of the PDR properties
obtained from such models toward W48A with those obtained from models which
take into account the photo and chemical processes in the PDR 
shows that detailed modeling gives values within the factors quoted above 
(Jeyakumar \etal\ 2007 in preparation). 

\section{\alfven speed in molecular clouds and non-thermal width
of carbon recombination lines}
\label{sec3}

Molecular clouds are largely neutral with a small admixture of 
partially ionized heavy elements with their free electrons 
(ionization fraction 10$^{-5}$). Any wave motion of the magnetic fields
in molecular clouds are strongly coupled to the ions. 
For those waves with $\lambda$  \gsim~$\lambda_{min}$, the time scale
of momentum transfer between ions and neutrals is smaller than
the magnetic perturbation time scale and hence neutrals
are also coupled to such waves (\nocite{am75}Arons \& Max 1975). For typical
magnetic fields of 1 mG near `dense' regions in star forming clouds 
and neutral densities of 10$^6$ \cmthree\  
(eg. \nocite{jmn89}Johnston, Migenes \& Norris 1989), 
$\lambda_{min} \sim 3 \times 10^{-7}$ pc, which is
much smaller than the size of these dense regions (fraction
of a parsec). Thus magnetic waves with $\lambda$  
\gsim~$3 \times 10^{-7}$ pc can exist in dense regions
and the characteristic speed (\alfven speed) of these
waves is determined by the total density (ie ion + neutral density) 
of the molecular cloud (\nocite{am75}Arons \& Max 1975). 

\begin{figure}
\includegraphics[width=3.0in, height=3.3in, angle = -90]{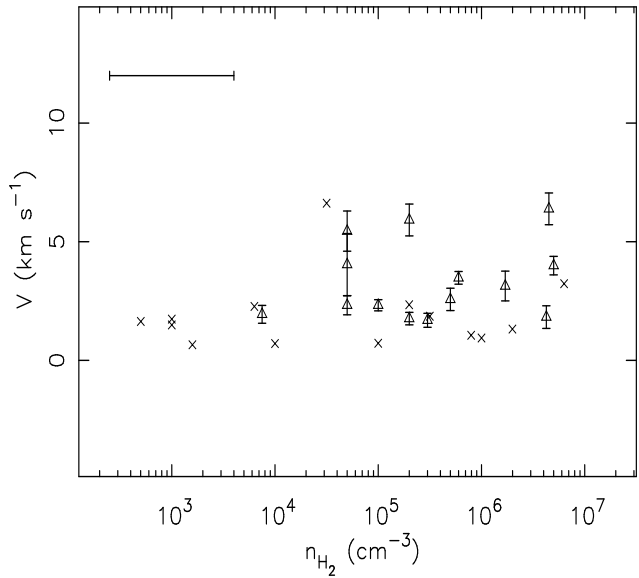}
\caption{\alfven speed in molecular clouds (crosses) and non-thermal 
velocity dispersion of carbon lines (triangles) observed toward PDRs
near \UCHII s are plotted against density in these regions.
\alfven speeds are estimated from the molecular line data compiled
by Crutcher (1999). Errors in the non-thermal velocity dispersion 
are derived from the uncertainties in the 
measurements and estimation of the model parameters.
The expected range of the estimated densities for the PDR is shown by
the horizontal bar. The plot shows that
the non-thermal component of the CRL width is ``constant''
for almost all PDRs and has a median value similar to that
of \alfven speed in molecular cloud.}
\label{fig2}
\end{figure}

\nocite{c99}Crutcher (1999) compiled the available sensitive Zeeman measurements
of magnetic field strengths in molecular clouds and their neutral densities. Using
this data, I
estimate the \alfven speed $V_A$ (units of cm s$^{-1}$) 
in these clouds;
\be 
V_A = \frac{2 B_{los}}{\sqrt{4\pi\mu n_H m_H}},
\label{eq:valph}
\ee
where $n_H$ is the hydrogen atom density in units of \cmthree,
$\mu = 1.4$ is the effective mass of an H+He gas with cosmic abundance 
and $m_H$ is the mass of the hydrogen atom in gm.  The measured
line-of-sight magnetic field, $B_{los}$, in units of G, 
is multiplied by 2 to convert
it into total field strength (\nocite{c99}Crutcher 1999). Fig.~\ref{fig2}
shows the estimated \alfven speed in the molecular cloud sample
taken from Crutcher (1999).  The estimated \alfven speeds are
confined to a narrow range
between 0.7 and 4 \kms\ with a median value of 1.6 \kms. This ``constancy'' of
\alfven speed was noted earlier and has been understood from models 
of magnetic confinement of molecular clouds 
(eg. \nocite{mp95}Moucschovias \& Psaltis 1995; \nocite{b2000}Basu 2000).

We now compare the non-thermal velocity dispersion of carbon lines, 
$\sigma_{NT}$, observed from
PDRs with \alfven speed in molecular clouds. The non-thermal width
is obtained by removing the
thermal contribution, estimated using the gas temperature
$T_{PDR}$ (see Table~\ref{tab1}), from the observed CRL width. 
The non-thermal line widths are plotted in Fig.~\ref{fig2} for the corresponding PDR 
densities inferred from carbon line modeling. 
The non-thermal widths have values
between 1.6 and 6 \kms with a median value of 2.9 \kms. 
Fig.~\ref{fig2} shows
that the non-thermal line widths are almost ``constant'' over three orders
of magnitude in density. The ``constancy'' of the non-thermal
line widths and their magnitudes are similar (less than a factor of 2) 
to those inferred for \alfven waves in molecular clouds.

The gas phase carbon is ionized in PDR and hence is  
strongly coupled to the magnetic field in these regions. 
If we consider similar physical parameters in PDRs as in
dense molecular regions ($n_H \sim 10^6$ \cmthree; $B \sim$ 1 mG)
then $\lambda_{min}$ of $\sim 3 \times 10^{-7}$ pc is  
at least two orders of magnitude smaller than the typical
line-of-sight thickness of the PDR (ie $L_{\parallel}$).
Thus \alfven waves with $\lambda$ \gsim~$\lambda_{min}$
exist in the PDR and here I consider their contribution to
the non-thermal width of the observed carbon lines. The amplitude of
the velocity of carbon ions, $\delta v$, due to these waves
is related to the magnetic perturbation amplitude 
$\delta B$ through the equation (eg. \nocite{am75}Arons \& Max 1975) 
\be
\delta v = \frac{\delta B}{\sqrt{4\pi\mu n_H m_H}}.
\ee
The observed non-thermal velocity dispersion is approximately
given by $\delta v$. It is usually assumed 
that $\delta B \sim B$, in that case, the right hand side of 
Eq. (2) becomes identical to that of Eq.~(\ref{eq:valph}). 
Based on these considerations and the 
characteristics exhibited by the non-thermal velocity dispersion
and \alfven speed in molecular clouds (see above) we assume that the
velocity dispersion of carbon lines is 
an estimate of the \alfven speed in the PDR. 

The observed velocity dispersion is usually scaled by $\sqrt{3}$
to convert it to 3D velocity dispersion. This scaling assumes random magnetic
field orientation along line-of-sight and random polarization
of \alfven waves. Carbon lines are observed from PDRs near \UCHII s
where shocks are present. In such shocked regions only the tangential
component of the magnetic field is amplified and the
\alfven speed associated with this component is scaled
by the square root of the density compression ratio
(\nocite{mz95}McKee \& Zweibel 1995). Hence the
scaling factor needed to convert the observed velocity dispersion
to a 3D dispersion is uncertain. Direct observation
of the Zeeman effect of CRLs may help in determining
this factor (see Section~\ref{sec5}). Here I note a systematically
high value (a factor of 1.8) for the CRL velocity dispersion
compared to the \alfven speed in molecular clouds, 
which may be an indication of higher \alfven speeds 
in PDR shocks. 

\section{Magnetic field in Photo-dissociation regions} 
\label{sec4}

The magnetic field strength is obtained using Eq~(\ref{eq:valph})
by substituting the estimated non-thermal velocity dispersion of CRL for
the \alfven speed and using the neutral density obtained 
from modeling the CRL emission
(eg. \nocite{retal05b}Roshi \etal\ 2005b). Field strength values
thus obtained are tabulated in Table.~\ref{tab1}. These values
represent the total magnetic field strength in the PDR.
Based on the expected range of the derived physical properties
of the PDR (see Section~\ref{sec2}), the estimated uncertainty
in $B$ is typically a factor of 2.5.

\begin{figure}
\includegraphics[width=2.5in, height=3.3in, angle = -90]{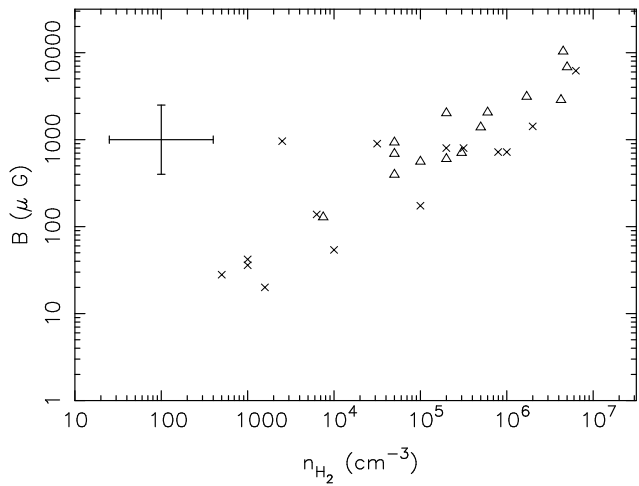}
\caption{Log-log plot of the magnetic field strength vs neutral density.
The plot compares the magnetic fields measured in molecular clouds
(crosses; data taken from Crutcher 1999) with those estimated 
using recombination line data (triangles). Comparison is 
done for similar $H_2$ densities in the 
molecular cloud and the PDR.  Such a comparison is possible since 
earlier observations toward molecular clouds show that the magnetic
field scales with density (Crutcher 1999).
The plot shows that the field strengths 
estimated using CRL data (see Section.~\ref{sec4})  
are comparable with those
measured in molecular clouds with densities similar to 
PDR densities.
The expected range of density
and estimated magnetic field are shown by horizontal and
vertical error bars respectively. }
\label{fig3}
\end{figure}

In Fig.~\ref{fig3}, we compare the estimated magnetic field with 
those measured toward molecular clouds. The ordinate of the plot
is the number density of $H_2$ molecules. Here, the estimated
field strength values in the PDR are compared with those measured 
in molecular
clouds with similar density. Such a comparison is possible since 
earlier observations toward molecular clouds show that the magnetic
field scales with density ($n_{H_2} \propto \rho^{0.47}$; 
\nocite{c99}Crutcher 1999).  To produce Fig.~\ref{fig3},
the neutral density of the PDR given in Table.~\ref{tab1} is divided by 2
to convert it into number density of H$_2$. As seen in the figure
the estimated magnetic field strengths in the PDR compare well within errors with
those measured toward molecular clouds with similar density.

Magnetic field measurements using Zeeman effect of CRLs near 1.4 GHz 
were attempted toward a few \HII\ regions (\nocite{s84}Silverglate 1984)
of which W48 and S88B are of interest here. The 1.4 GHz CRL 
emission toward W48 does not originate from the PDR associated
with the \UCHII\ (\nocite{retal05a}Roshi \etal\ 2005a) and
hence a comparison of the upper limit on the field strength obtained by 
\nocite{s84}Silverglate (1984) with the estimated value here is
not meaningful. Toward S88B, the upper limit for the magnetic 
field strength obtained is consistent with estimated
values given in Table~\ref{tab1}.  Magnetic field
measurements using OH or \HI\ Zeeman effects are available 
in literature toward a few \UCHII s listed in Table~\ref{tab1}.
I compare the field strength given in Table~\ref{tab1} 
with the results of these observations. Note that OH and \HI\   
lines may originate from different
spatial locations compared to the CRL forming region. 
From OH Zeeman observations toward S88B a field strength
in the range 0.1 -- 0.3 mG was obtained by Sarma \etal\ (2006 in preparation)
consistent with our estimate. 
\nocite{vg90}Van der Werf \& Goss (1990) using \HI\ Zeeman observations
measured a peak magnetic field of 0.1 mG in the $-$45 \kms
component observed toward W3. The CRL LSR velocity is comparable with this velocity
and the field strength is consistent with the estimated value.
\nocite{bt01}Brogan \& Troland (2001) measured a maximum
field strength of 0.3 mG toward W49A. They used Zeeman effect
of \HI\ to measure the field strength with an angular resolution
of 25\arcsec. The measured value is about twenty times smaller than 
the estimated value; possibly because the two tracers (\HI\ and CRL) 
do not probe the same region in this case.

\section{Discussion}
\label{sec5}

As mentioned in Section~\ref{sec2}, CRL emission is dominated by stimulated 
emission for the transitions listed in Table~\ref{tab1}.
Because of the dominance of stimulated emission, carbon line 
is detected only from the near side of the \UCHII. The non-detection 
from far side of \UCHII\  means that the width of CRL is not contributed
to by the expansion of the \UCHII\ (if the \UCHII\ were expanding).
Thus the line width has contribution only from thermal and non-thermal motions.    
The carbon line width may also have contributions from large scale motions such
as outflows. However, the similarity of line widths observed from different sources
(see Table~\ref{tab1}) may indicate that this contribution is small.
Thus the field strength estimated in Section~\ref{sec4} may not
be affected by such large scale motions. However, presence of 
any non-magnetic turbulence would affect the estimation of the field strength and
hence the values obtained should be considered as upper limits.


The magnetic origin of CRL width can be confirmed by 
measuring the field strength using Zeeman effect of carbon lines 
from the PDRs and comparing it with the
values estimated in Section~\ref{sec4}. If confirmed then 
multi-frequency CRL observations form another tool to deduce 
the strength of magnetic fields near star forming regions.

\acknowledgments

I acknowledge many useful discussions with K. Subramanian, S. Sridhar, 
A. A. Deshpande, D. Bhattacharya and helpful comments from an anonymous referee.

\clearpage

\clearpage


\clearpage

\clearpage

\begin{table}
\small
\begin{center}
\caption{Carbon recombination line data. \label{tab1}}
\begin{tabular}{lrrccrrcc} 
\tableline\tableline
Source & Transition & $\Delta V$ &  $T_{PDR}$  & $n_{C^+}$ & $n_H $ & $L_{\parallel}$ & $B$ & Ref \\ 
name &    & (\kms) &  (K) & (\cmthree) & ($\times$ 10$^{6}$ \cmthree) & ($\times$ 10$^{-3}$ pc) & (mG) & \\ 
\tableline 
W48A        &  C76$\alpha$ &  4.5(0.8) &  500 & 2500 & 8.5 &  0.3 &  2.9 &  1 \\
W49G\tablenotemark{a} &  C91$\alpha$ &  9.5(0.8) &  500 & 3000 &10.0 &  0.1 & 6.8   &  2 \\
W49J\tablenotemark{a} &  C91$\alpha$ & 15.1(1.5) &  500 & 2800 & 9.0 &  0.4 & 10.4  &  2 \\
S88B-east   &  C92$\alpha$ &  4.4(0.3) &  600 &   80 & 0.4 &160.0 &  0.6 &  3\\
S88B-west   &  C92$\alpha$ &  5.6(0.4) &  400 &   40 & 0.2 &220.0 &  0.6 &  3   \\
GGD12$-$15    &  C92$\alpha$ &  4.7(0.7) &  330 &      & 0.02&130.0 &  0.1 & 4\\
W3A         & C168$\alpha$ &  5.5(0.9) &  100 &      & 0.1 & 50 & 0.4 & 5 \\
G32.80+0.19\tablenotemark{b} &  C89$\alpha$ &  7.5(1.3) &  500 & 1000 & 3.4 &  1.1 & 3.1 &  6 \\
G37.87$-$0.40\tablenotemark{b} &  C89$\alpha$ & 14.0(1.5) &  500 &  120 & 0.4 & 29.0 & 2.0 &  6 \\
G43.24$-$0.05\tablenotemark{b} &  C89$\alpha$ &  6.2(0.9) &  500 &  300 & 1.0 &  9.0 & 1.4 &  6\\
G45.12+0.13\tablenotemark{b} &  C89$\alpha$ & 12.9(1.9) &  500 &   40 & 0.1 &124.0 & 0.9 &  6 \\
G45.45+0.0\tablenotemark{b} &  C89$\alpha$ &  9.6(2.9) &  500 &   30 & 0.1 &150.0 & 0.7 &  6\\
G70.29+1.60\tablenotemark{b} &  C89$\alpha$ &  8.3(0.5) &  500 &  350 & 1.2 &  8.6 & 2.1 &  6\\
G70.33+1.59\tablenotemark{b} &  C89$\alpha$ &  4.2(0.4) &  500 &  170 & 0.6 & 25.0 & 0.7 &  6 \\ 
\tableline
\end{tabular}
\tablenotetext{a}{Modeling of carbon line emission for these sources has provided upper limits 
for $n_{C^+}$, $n_H$ and lower limits for $L_{\parallel}$.}
\tablenotetext{b}{The data for these sources are taken from the Arecibo survey. Modeling of carbon line emission for these sources has provided lower limits for $n_{C^+}$, $n_H$ and upper limits
for $L_{\parallel}$.}
  
\tablenotetext{}{References: (1)Roshi \etal\ (2005a); (2) Roshi \etal\ (2006); 
(3) Garay \etal\ (1998); (4) Gomez \etal\ (1998); (5) Kantharia \etal\ (1998); (6) Roshi \etal\ (2005b). }
\end{center}
\end{table}

\end{document}